# Relationship between the $T_C$ of smart meta-superconductor Bi(Pb)SrCaCuO and inhomogeneous phase content

Honggang Chen, Mingzhong Wang, Yao Qi, Yongbo Li and Xiaopeng Zhao*

Smart Materials Laboratory, Department of Applied Physics,

Northwestern Polytechnical University, Xi'an 710129, China;

*Correspondence: Prof. Xiaopeng Zhao, E-mail: xpzhao@nwpu.edu.cn

**Abstract:** A smart meta-superconductor Bi(Pb)SrCaCuO (B(P)SCCO) may increase the critical transition temperature ($T_C$) of B(P)SCCO by electroluminescence (EL) energy injection of inhomogeneous phases. However, the increase amplitude $\Delta T_C$ ($\Delta T_C = T_C - T_{C,pure}$) of $T_C$ is relatively small. In this study, a smart meta-superconductor B(P)SCCO with different matrix sizes was designed. Three kinds of raw materials with different particle sizes were used, and different series of $Y_2O_3:Sm^{3+}$, $Y_2O_3$, $Y_2O_3:Eu^{3+}$, and $Y_2O_3:Eu^{3+}$+Ag doped samples and pure B(P)SCCO were prepared. Results indicated that the $T_C$ of the $Y_2O_3$ or $Y_2O_3:Sm^{3+}$ non-luminescent dopant doping sample is lower than that of pure B(P)SCCO. However, the $T_C$ of the $Y_2O_3:Eu^{3+}$+Ag or $Y_2O_3:Eu^{3+}$ luminescent inhomogeneous phase doping sample is higher than that of pure B(P)SCCO. With the decrease of the raw material particle size from 30 to 5 μm, the particle size of the B(P)SCCO superconducting matrix in the prepared samples decreases, and the doping content of the $Y_2O_3:Eu^{3+}$+Ag or $Y_2O_3:Eu^{3+}$ increases from 0.2% to 0.4%. Meanwhile, the increase of the inhomogeneous phase content enhances the $\Delta T_C$. When the particle size of raw material is 5 μm, the doping concentration of the luminescent inhomogeneous phase can be increased to 0.4%. At this time, the zero-resistance temperature and onset transition temperature of the $Y_2O_3:Eu^{3+}$+Ag doped sample are 4 and 6.3 K higher than those of pure B(P)SCCO, respectively.

## 1. Introduction

Since the discovery of superconductivity, raising the critical transition temperature ($T_C$) of superconductivity to obtain a room-temperature superconductor has been the main goal of superconductivity research. In 2011, Cavalleri et al. used a mid-infrared femtosecond laser pulse to induce the transformation of $La_{1.675}Eu_{0.2}Sr_{0.125}CuO_4$ from a non-superconducting phase to a transient superconducting phase [1]. Subsequently, they used similar experimental methods to induce transient superconducting transitions in $La_{1.84}Sr_{0.16}CuO_4$, $YBa_2Cu_3O_{6.5}$, and $K_3C_{60}$ [2-4]. Since then, the use of light to change the superconducting properties of materials has been gradually recognized by researchers. In 2015, German researchers observed the superconductivity of 203 K in a sulfur hydride system at 155 Gpa [5]. And they obtained a superconductivity of 250 K at 170 GPa for $LaH_{10}$ in 2019 [6]. In 2020, Dias et al. achieved a superconducting transition of 287.7 K at approximately 267 GPa in a carbonaceous sulfur hydride system [7]. Although the $T_C$ continues to refresh, the ultra-high pressure, extremely small sample size, and material instability under ambient pressure limit further applications.

The known stable high-temperature superconductors are all layered copper oxides with a perovskite-like structure. Among them, Bi-based superconductors (BSCCO) with a $T_C$ beyond 100 K have become the most promising material for scientific and industrial applications due to their several advantages, such as high $T_C$, low oxygen sensitivity, and absence of rare earth [8-10]. Bi-based superconductors have three superconducting phases of Bi2201 ($T_C$=20 K), Bi2212 ($T_C$=85 K), and Bi2223 ($T_C$=110 K). They are symbiotic with one another, and a pure single phase is difficult to obtain [11-14]. By partial substitution of Bi by Pb, the volume fraction of the high-temperature phase Bi2223 can be increased, thus making the Bi2223 phase easier to synthesize and increasing its stability [15, 16]. Although Bi-based superconductors are called high-temperature superconductors, a large gap still exists between their $T_C$ and practical applications. Hence, researchers have done substantial work to improve their $T_C$. At present, the most commonly used method is chemical doping, such as doping Cs [17], Al [18], $SnO_2$ [19], $ZrO_2$ [20], $Ca_2B_2O_5$ [21], and others in BSCCO. However, most of the dopants are unstable at high temperatures and will react with superconductors, the results are not ideal.

Metamaterial is a kind of composite material with an artificial structure [22-24]. Smolyaninov et al. proposed that a metamaterial superconductor with an effective dielectric constant that is less or approximately equal to zero can exhibit a higher $T_C$ and confirmed their idea in the experiment [25-27]. In 2007, our group doped inorganic EL material in a superconductor to increase $T_C$ via EL. Jiang et al. [28] introduced a ZnO EL material into B(P)SCCO superconductors for the first time. The results showed that the $T_C$ of B(P)SCCO was slightly reduced due to the low EL intensity and the high doping concentration of ZnO. In recent years, we constructed a smart meta-superconductor $MgB_2$. The conventional superconductor $MgB_2$ is doped with $Y_2O_3$:$Eu^{3+}$, $Y_2O_3$:$Eu^{3+}$+Ag sheets, and $Y_2O_3$:$Eu^{3+}$ rods [29-31] of different sizes. The results showed that the doping of $Y_2O_3$:$Eu^{3+}$ EL materials increases the $T_C$ of $MgB_2$. We believe that the $Y_2O_3$:$Eu^{3+}$ materials generate an EL under the action of an external electric field and that EL energy injection promotes the formation of Cooper pairs, the $T_C$ of $MgB_2$ is improved via EL [32-36]. At the same time, a smart meta-superconductor B(P)SCCO consists of the B(P)SCCO matrix and an inhomogeneous phase $Y_2O_3$:$Eu^{3+}$ or $Y_2O_3$:$Eu^{3+}$+Ag was constructed in BSCCO superconductors. It was showed that the addition of $Y_2O_3$:$Eu^{3+}$ and $Y_2O_3$:$Eu^{3+}$+Ag inhomogeneous phase increases the $T_C$ of B(P)SCCO [37]. However, the $T_C$ of the pure B(P)SCCO prepared is low, the increase amplitude $\Delta T_C$ ($\Delta T_C = T_C - T_{C,pure}$) is small, and the influencing factors are not clear.

On the basis of previous research, a smart meta-superconductor consists of an inhomogeneous phase and the B(P)SCCO matrix with different sizes is designed in this study. Three kinds of raw materials with different particle sizes are used, and three different series of samples are prepared by solid-state sintering. The transition width and $T_C$ of the prepared samples are consistent with the literature [38-40]. The $T_C$ of $Y_2O_3$:$Sm^{3+}$ and $Y_2O_3$ non-luminescent dopant doping sample is lower than that of pure B(P)SCCO ($\Delta T_C < 0$). However, the $T_C$ of $Y_2O_3$:$Eu^{3+}$ and $Y_2O_3$:$Eu^{3+}$+Ag luminescent inhomogeneous phases doping sample is higher than that of pure B(P)SCCO ($\Delta T_C > 0$). With the decrease of the raw material particle size, the particle size of the B(P)SCCO superconducting matrix in the prepared samples decreases, the doping content of the $Y_2O_3$:$Eu^{3+}$+Ag and $Y_2O_3$:$Eu^{3+}$ inhomogeneous phases can increase. Meanwhile, the increase of luminescent inhomogeneous phases content amplifies the increase amplitude $\Delta T_C$ of $T_C$.

## 2. Model

On the basis of the idea of metamaterials, a smart meta-superconductor B(P)SCCO model is constructed, as shown in Figure 1. The model is composed of B(P)SCCO superconducting particles of the matrix material and a $Y_2O_3$:$Eu^{3+}$+Ag or $Y_2O_3$:$Eu^{3+}$ luminescent inhomogeneous phase. The gray hexagons in the figure represent B(P)SCCO superconducting particles. The superconducting particles are composed of numerous small crystal grains, and the black lines in the gray hexagons represent the B(P)SCCO grain boundaries. The luminescent inhomogeneous phase is located around the B(P)SCCO superconducting particles and is represented by a discontinuous white color. When measuring the $R$-$T$ curve of the samples, under the action of an external electric field, the B(P)SCCO superconducting particles act as microelectrodes, which excite the EL of the luminescent inhomogeneous phase. Moreover, the formation of electron pairs can be promoted via EL energy injection, and the $T_C$ of B(P)SCCO will be improved. By adjusting the external electric field to control the EL and adjust the $T_C$, a smart meta-superconductor is realized.

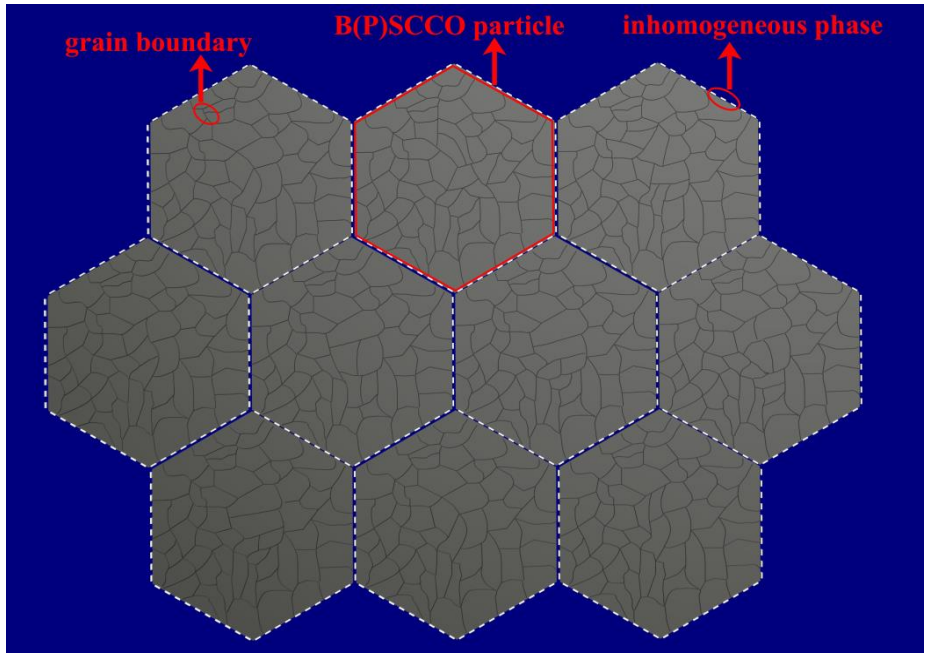

**Figure 1.** The model of the smart meta-superconductor B(P)SCCO.

By changing the size of the matrix material in the smart meta-superconductor B(P)SCCO model, more luminescent inhomogeneous phases can be accommodated in the same space volume as the size of the B(P)SCCO superconducting particle decreases. In addition, the distribution of the inhomogeneous phases is more uniform. Hence, the formation of electron pairs is enhanced, and the increase amplitude $\Delta T_C$ of $T_C$ will be further increased.

## 3. Experiment

### 3.1. Preparation of luminescent inhomogeneous phases and non-luminescent dopants

The preparation process and related characterization of the $Y_2O_3$:$Eu^{3+}$+Ag topological luminophore inhomogeneous phase were described in Ref. [30, 37]. The same preparation process is used by changing the raw materials to obtain $Y_2O_3$:$Eu^{3+}$ luminescent inhomogeneous phase and $Y_2O_3$ and $Y_2O_3$:$Sm^{3+}$ non-luminescent dopants.

### 3.2. Preparation of pure B(P)SCCO superconductors

A certain amount of raw materials according to the molar ratio $Bi_2O_3:PbO:SrCO_3:CaCO_3:CuO=0.85:0.4:2.00:2.00:3.00$ was weighed and placed in an agate tank for 500 r/min ball milling for 20, 50, or 80 hours. The raw material of the 20-h ball milling was dried and sieved with a 500-mesh stainless steel sieve to obtain B(P)SCCO raw material S1 with a particle size of approximately 30 μm. The raw material after 50-h ball milling was filtered and dried to obtain B(P)SCCO raw material S2 with a particle size of approximately 15 μm. Meanwhile, the raw material of 80-h ball milling was filtered and dried to obtain B(P)SCCO raw material S3 with a particle size of approximately 5 μm. Then, raw materials S1, S2, and S3 were kept at 840 ℃ for 50+50 h, respectively. They were fully ground once in the middle to obtain three kinds of B(P)SCCO calcined powder. The calcined powder was sufficiently ground and pressed into a pellet of 12 mm diameter and 2 mm thickness. The pellet was kept at 840 ℃ in the air for 30+120 h. It was milled and pressed again in the middle to obtain a pure B(P)SCCO sample.

### 3.3. Preparation of doping B(P)SCCO superconducting samples

A certain amount of $Y_2O_3:Eu^{3+}$+Ag and $Y_2O_3:Eu^{3+}$ luminescent inhomogeneous phases or $Y_2O_3$, $Y_2O_3:Sm^{3+}$ dopants was fully mixed with three kinds of B(P)SCCO calcined powder. Then, the mixture was pressed into a pellet. The pellet was kept at 840 ℃ in the air for 30+120 h. It was milled and pressed again in the middle to obtain samples doped with different dopants.

### 3.4. Characterization

X-ray diffraction (XRD) patterns within the range of $3°≤2θ≤60°$ were obtained using an Hitachi XRD-7000 diffractometer with Cu Kα radiation at a scanning rate of 0.1 °/s. A FEI Verios G4 scanning electron microscope (SEM) was used to analyze the microstructure of the sintered samples. Resistivity versus temperature measurements were performed on each of the samples using the standard four-probe technique in a liquid helium cryogenic system. In addition, 0.1 mA current was applied. Keithley digital nanovoltmeter was used to measure the high resolution voltage across the sample.

In the experiment, the particle size of the raw materials S1, S2, and S3 obtained by ball milling and drying decreased in turn. The particle size of the raw materials S1, S2, and S3 were approximately 30, 15, and 5 μm, respectively. The samples prepared with raw materials S1, S2, and S3 were recorded as A series samples (as shown in Table 1), B series samples (as shown in Table 2), and C series samples (as shown in Table 3), respectively.

**Table 1.** Dopant, doping concentration, $T_{C,0}$, and $T_{C,on}$ of A series samples prepared from raw material S1 (30 μm).

| Sample | Dopant | Doping concentration (wt%) | $T_{C,0}$/K | $T_{C,on}$/K | $ΔT_{C,0}$/K | $ΔT_{C,on}$/K |
|---|---|---|---|---|---|---|
| A1 | None | 0 | 105 | 113.7 | 0 | 0 |
| A2 | $Y_2O_3:Sm^{3+}$ | 0.2 | 102.5 | 112.7 | -2.5 | -1 |
| A3 | $Y_2O_3$ | 0.2 | 103.5 | 113.1 | -1.5 | -0.6 |
| A4 | $Y_2O_3:Eu^{3+}$ | 0.2 | 105.5 | 114.4 | 0.5 | 0.7 |
| A5 | $Y_2O_3:Eu^{3+}$+Ag | 0.2 | 106 | 114.8 | 1 | 1.1 |
| A6 | $Y_2O_3:Eu^{3+}$+Ag | 0.3 | 105.5 | 114 | 0.5 | 0.3 |

**Table 2.** Dopant, doping concentration, $T_{C,0}$, and $T_{C,on}$ of B series samples prepared from raw material S2 (15 μm).

| Sample | Dopant | Doping concentration (wt%) | $T_{C,0}$/K | $T_{C,on}$/K | $\Delta T_{C,0}$/K | $\Delta T_{C,on}$/K |
|---|---|---|---|---|---|---|
| B1 | None | 0 | 100.5 | 107.9 | 0 | 0 |
| B2 | $Y_2O_3$:$Sm^{3+}$ | 0.2 | 97.5 | 107 | -3 | -0.9 |
| B3 | $Y_2O_3$ | 0.2 | 98 | 106.4 | -2.5 | -1.5 |
| B4 | $Y_2O_3$:$Eu^{3+}$ | 0.2 | 101.5 | 110.4 | 1 | 2.5 |
| B5 | $Y_2O_3$:$Eu^{3+}$+Ag | 0.2 | 102.5 | 110.8 | 2 | 2.9 |
| B6 | $Y_2O_3$:$Eu^{3+}$+Ag | 0.3 | 103.5 | 112.2 | 3 | 4.3 |

**Table 3.** Dopant, doping concentration, $T_{C,0}$, and $T_{C,on}$ of C series samples prepared from raw material S3 (5 μm).

| Sample | Dopant | Doping concentration (wt%) | $T_{C,0}$/K | $T_{C,on}$/K | $\Delta T_{C,0}$/K | $\Delta T_{C,on}$/K |
|---|---|---|---|---|---|---|
| C1 | None | 0 | 100 | 107.6 | 0 | 0 |
| C2 | $Y_2O_3$:$Sm^{3+}$ | 0.3 | 96.5 | 106.6 | -3.5 | -1 |
| C3 | $Y_2O_3$ | 0.3 | 97 | 106 | -3 | -1.6 |
| C4 | $Y_2O_3$:$Eu^{3+}$ | 0.3 | 101 | 110.4 | 1 | 2.8 |
| C5 | $Y_2O_3$:$Eu^{3+}$+Ag | 0.3 | 103.5 | 112.1 | 3.5 | 4.5 |
| C6 | $Y_2O_3$:$Eu^{3+}$ | 0.4 | 102.5 | 112.2 | 2.5 | 4.6 |
| C7 | $Y_2O_3$:$Eu^{3+}$+Ag | 0.4 | 104 | 113.9 | 4 | 6.3 |

## 4. Results and discussion

Figure 2(a) shows the XRD patterns of pure B(P)SCCO A1, B1, and C1 prepared by the solid-state sintering of raw materials S1, S2, and S3. The peak intensities and positions of the diffraction pattern demonstrate that the main phase of prepared pure B(P)SCCO is the high-temperature phase Bi2223 (the volume fraction is approximately 95%). A small amount of the low-temperature phase Bi2212 is detected, and no other impurity phases exist. Moreover, the particle size of the raw material does not affect the phase formation of prepared samples. The high-temperature phase Bi2223 and the low-temperature phase Bi2212 are marked with a circle and a square, respectively. The calculation formula for the relative volume fraction of the high-temperature phase Bi2223 and the low-temperature phase Bi2212 of all samples is as follows [41, 42]:

$$\text{Bi2223}(\%) \approx \frac{\sum I(Bi2223)}{\sum I(Bi2223)+\sum I(Bi2212)} \times 100\%,$$

$$\text{Bi2212}(\%) \approx \frac{\sum I(Bi2212)}{\sum I(Bi2223)+\sum I(Bi2212)} \times 100\%,$$

where $I$ is the intensity of the Bi2223 phase and the Bi2212 phase in the XRD pattern. Figure 2(b) depicts the XRD patterns of doped samples (A1, A2, A3, A4, A5, and A6) prepared by raw material S1. The addition of the $Y_2O_3$:$Sm^{3+}$ and $Y_2O_3$ non-luminescent dopants or the $Y_2O_3:Eu^{3+}$ and $Y_2O_3:Eu^{3+}$+Ag luminescent inhomogeneous phases does not introduce other impurity phases.

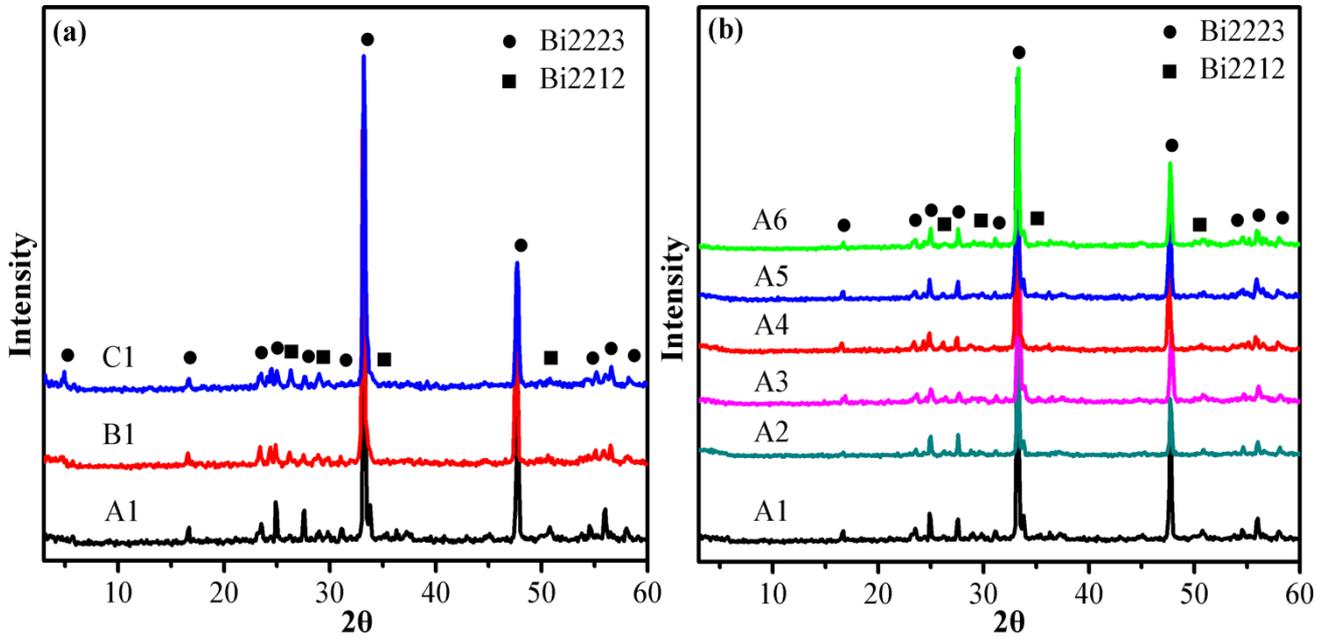

**Figure 2.** XRD patterns of samples. **(a)** XRD patterns of pure B(P)SCCO A1, B1, and C1 prepared by the raw materials S1, S2, and S3. **(b)** XRD patterns of pure B(P)SCCO (A1) and B(P)SCCO doped with 0.2 wt% $Y_2O_3:Sm^{3+}$ (A2), 0.2 wt% $Y_2O_3$ (A3), 0.2 wt% $Y_2O_3:Eu^{3+}$ (A4), 0.2 wt% $Y_2O_3:Eu^{3+}$+Ag (A5), and 0.3 wt% $Y_2O_3:Eu^{3+}$+Ag (A6).

The microstructure is one of the important properties of high-temperature superconductors. Figures 3(a), 3(b), and 3(c) demonstrate the SEM images of pure B(P)SCCO A1, B1, and C1 prepared from raw materials S1, S2, and S3, respectively. In all samples, the dominant structures are plate-like structures randomly distributed due to the presence of the pores. As the size of the raw materials S1, S2, and S3 decreases, the size of the plate-like structures in the prepared sample diminishes. Meanwhile, the pores increase, thus being able to accommodate more $Y_2O_3:Eu^{3+}$+Ag and $Y_2O_3:Eu^{3+}$ inhomogeneous phases in the same space volume. In addition, the inhomogeneous phase distribution around the B(P)SCCO superconducting matrix is more uniform. Figure 3 (d) shows the SEM image of B(P)SCCO doped with 0.4 wt% $Y_2O_3:Eu^{3+}$+Ag (C7) prepared from the raw material S3. The doping of $Y_2O_3:Eu^{3+}$+Ag inhomogeneous phase does not affect the microstructure of B(P)SCCO.

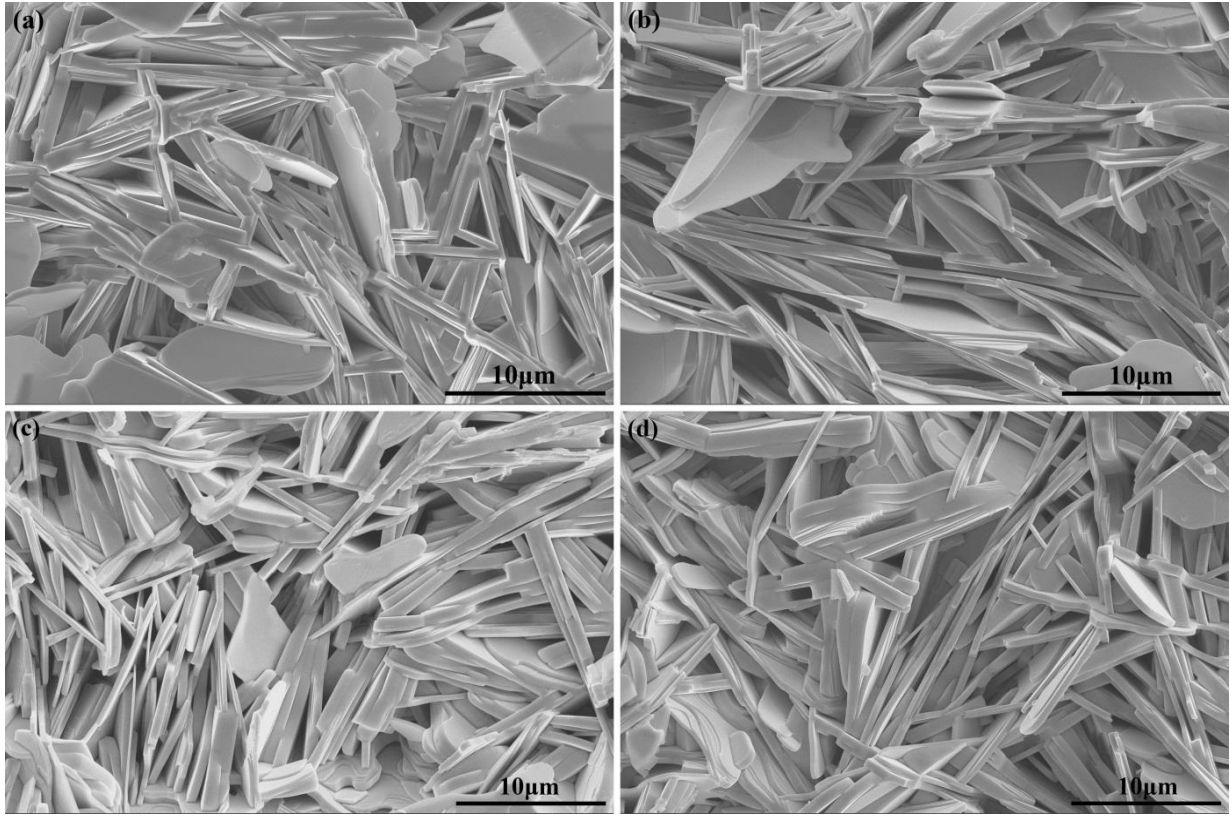

**Figure 3.** SEM images of samples. **(a), (b), (c)** SEM images of pure B(P)SCCO A1, B1, and C1 prepared from the raw materials S1, S2, and S3. **(d)** SEM image of B(P)SCCO doped with 0.4 wt% $Y_2O_3$:$Eu^{3+}$+Ag (C7) prepared from the raw material S3.

The standard four-probe method is used to test the *R-T* curve of the prepared samples from 50 K to room temperature. In the normal state region, all samples show the linear temperature dependence characteristic of Cu-oxide-based high-temperature superconductors and a superconducting transition between 90 and 115 K. Figure 4(a) shows the normalized *R-T* curve (the black curve) of the pure B(P)SCCO (A1) prepared from raw material S1. Through the relationship between temperature and normalized *dρ/dT* (the blue curve in Figure 4(a)), zero-resistance temperature $T_{C,0}$ and onset transition temperature $T_{C,on}$ can be obtained. Zero-resistance temperature $T_{C,0}$ is the temperature at which the resistance has just completely dropped to zero during the cooling process. Meanwhile, onset transition temperature $T_{C,on}$ is the temperature at which the *R–T* curve deviates from linear behavior. The $T_{C,0}$ and $T_{C,on}$ of the pure B(P)SCCO (A1) are 105 and 113.7 K, respectively, and the transition width is 8.7 K, which is consistent with the literature [38-40]. Figure 4(b) presents the normalized *R-T* curve of pure B(P)SCCO (A1) and B(P)SCCO doped with 0.2 wt% $Y_2O_3$:$Sm^{3+}$ (A2), $Y_2O_3$ (A3), $Y_2O_3$:$Eu^{3+}$ (A4), $Y_2O_3$:$Eu^{3+}$+Ag (A5), and 0.3 wt% $Y_2O_3$:$Eu^{3+}$+Ag (A6). Figure 4(c) and 4(d) present the $\Delta T_{C,0}$ and $\Delta T_{C,on}$ of A2, A3, A4, A5, and A6 relative to the pure B(P)SCCO (A1), respectively. The specific values of the $T_C$ are shown in Table 1. From the chart, we can observe that the $Y_2O_3$:$Sm^{3+}$ and $Y_2O_3$ non-luminescent dopants doping make the $T_C$ of B(P)SCCO lower ($\Delta T_C<0$). However, the addition of the $Y_2O_3$:$Eu^{3+}$+Ag and $Y_2O_3$:$Eu^{3+}$ inhomogeneous phases increase $T_C$ ($\Delta T_C>0$), which may be due to the luminescent inhomogeneous phase distributed around the B(P)SCCO superconducting particles to form a metamaterial structure with a special response. When measuring the *R-T* curve of the sample, B(P)SCCO superconducting particles act as the microelectrode, which excites the EL of the luminescent inhomogeneous phase. In addition, the formation of electron pairs is promoted via EL energy injection so that the $T_C$ of B(P)SCCO is improved. And the $T_{C,0}$ and $T_{C,on}$ of 0.2 wt% $Y_2O_3$:$Eu^{3+}$+Ag doped sample (A5) are higher than 0.3% $Y_2O_3$:$Eu^{3+}$+Ag doped sample (A6).

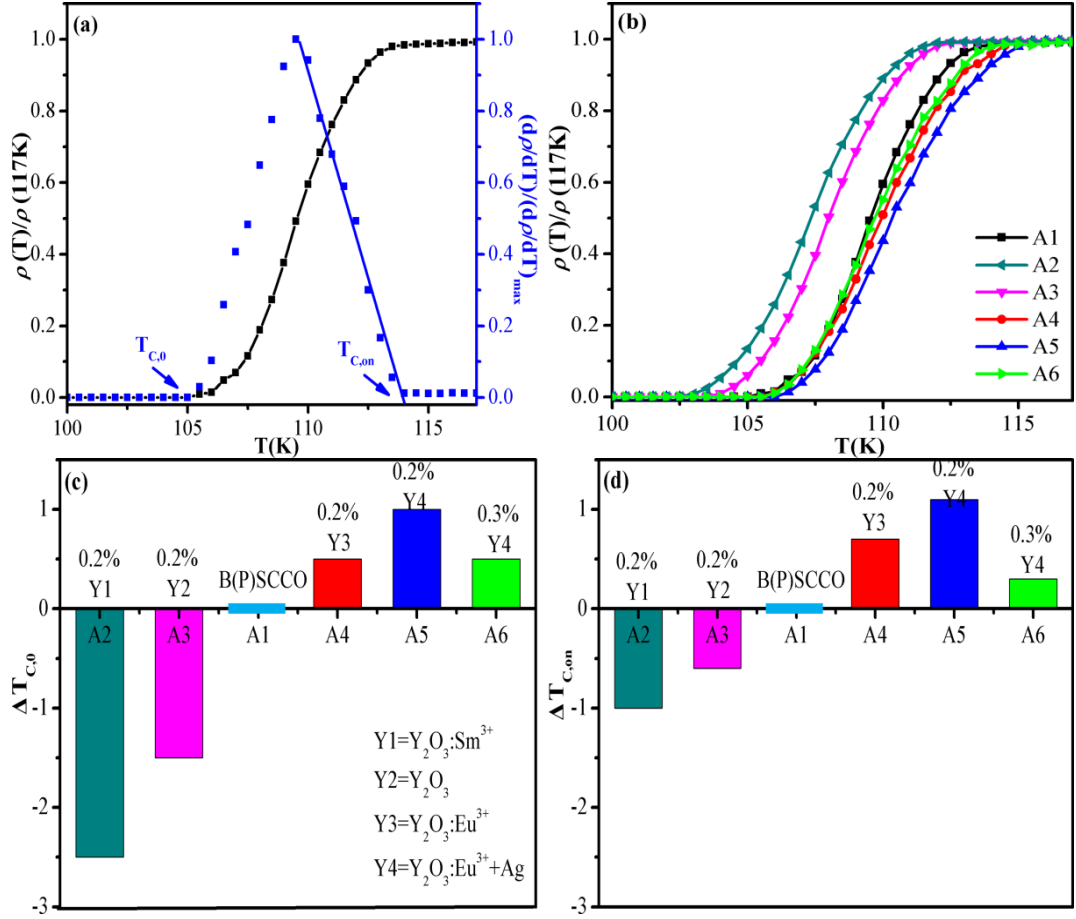

**Figure 4.** Superconductivity of A series samples. **(a)** Temperature dependence of normalized $d\rho/dT$ of pure B(P)SCCO (A1). **(b)** Temperature-dependent normalized resistivity curves of pure B(P)SCCO (A1) and B(P)SCCO doped with 0.2 wt% $Y_2O_3$:$Sm^{3+}$ (A2), 0.2 wt% $Y_2O_3$ (A3), 0.2 wt% $Y_2O_3$:$Eu^{3+}$ (A4), 0.2 wt% $Y_2O_3$:$Eu^{3+}$+Ag (A5), and 0.3 wt% $Y_2O_3$:$Eu^{3+}$+Ag (A6). **(c), (d)** The $\Delta T_{C,0}$ and $\Delta T_{C,on}$ of A2, A3, A4, A5, and A6 relative to A1.

Figure 5(a) demonstrates the normalized *R-T* curve of pure B(P)SCCO (B1) and 0.2 wt% $Y_2O_3$:$Sm^{3+}$ (B2), $Y_2O_3$ (B3), $Y_2O_3$:$Eu^{3+}$ (B4), $Y_2O_3$:$Eu^{3+}$+Ag (B5), and 0.3 wt% $Y_2O_3$:$Eu^{3+}$+Ag (B6) doped samples. Figures 5(b) and 5(c) represent increase amplitudes $\Delta T_{C,0}$ and $\Delta T_{C,on}$ of B2, B3, B4, B5, and B6 relative to the pure B(P)SCCO (B1), respectively. Figure 5(d) shows the normalized *R-T* curve of pure B(P)SCCO (C1) and B(P)SCCO doped with 0.3 wt% $Y_2O_3$:$Sm^{3+}$ (C2), 0.3 wt% $Y_2O_3$ (C3), 0.3 wt% $Y_2O_3$:$Eu^{3+}$ (C4), 0.3 wt% $Y_2O_3$:$Eu^{3+}$+Ag (C5), 0.4 wt% $Y_2O_3$:$Eu^{3+}$ (C6), and 0.4 wt% $Y_2O_3$:$Eu^{3+}$+Ag (C7). Figures 5(e) and 5(f) represent increase amplitudes $\Delta T_{C,0}$ and $\Delta T_{C,on}$ of C2, C3, C4, C5, C6, and C7 relative to the pure B(P)SCCO (C1), respectively. The $T_C$ of each sample is shown in Tables 2 and 3. The non-luminescent dopants $Y_2O_3$ and $Y_2O_3$:$Sm^{3+}$ doping make the $T_{C,0}$ and $T_{C,on}$ of B(P)SCCO decrease by 2-4 and 1-2 K, respectively. And as the doping content of the non-luminescent dopants increases, the $T_C$ of B(P)SCCO decreases more. While, the $Y_2O_3$:$Eu^{3+}$ and $Y_2O_3$:$Eu^{3+}$+Ag luminescent inhomogeneous phases doped sample have a higher $T_{C,0}$ and $T_{C,on}$ than the pure sample. $T_{C,on}$ increases more. And the $T_C$ of $Y_2O_3$:$Eu^{3+}$+Ag doping is higher than that of $Y_2O_3$:$Eu^{3+}$ doping. In addition, within a certain content range, the $\Delta T_{C,0}$ and $\Delta T_{C,on}$ increases as the content of the luminescent inhomogeneous phase increases.

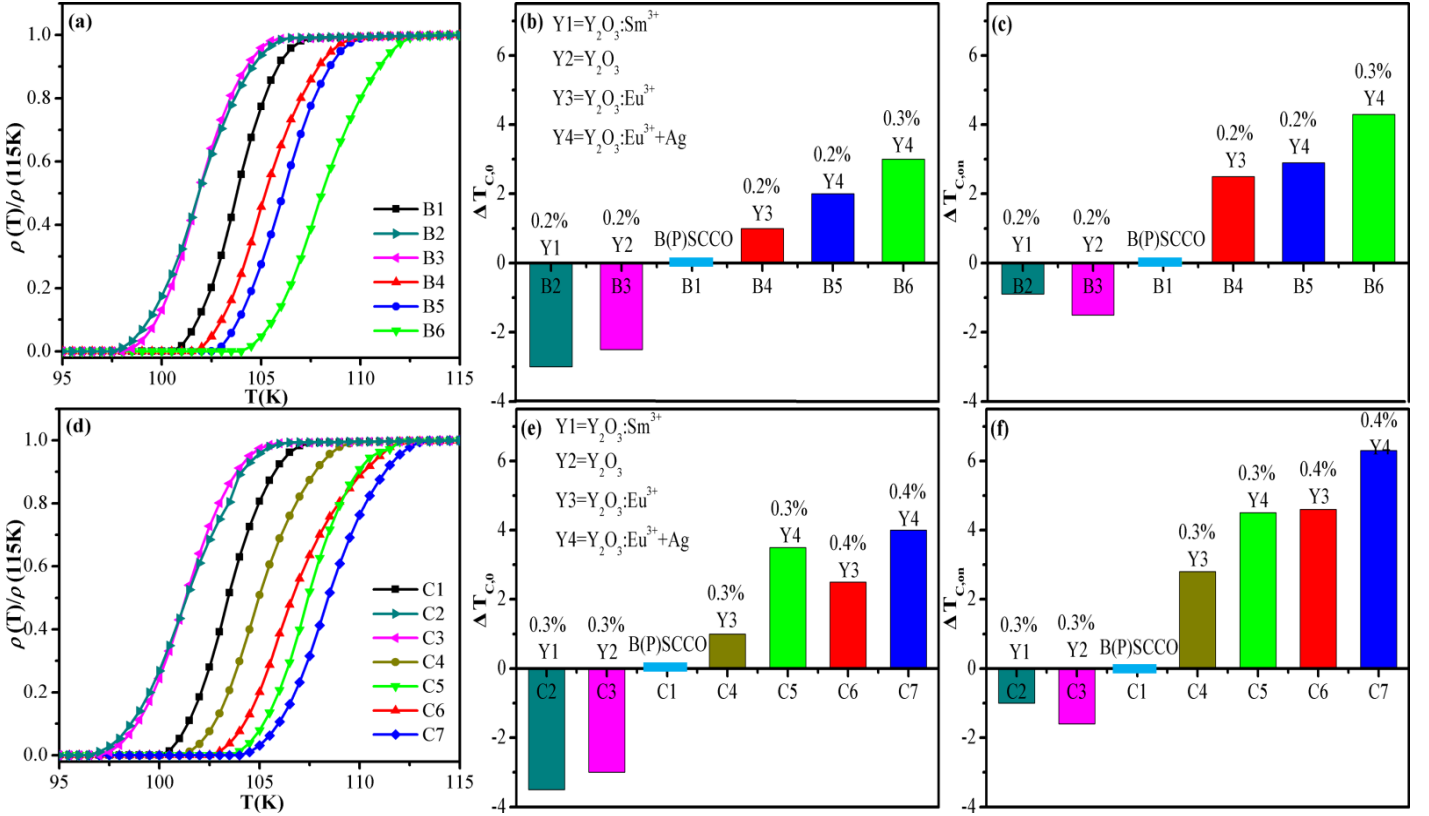

**Figure 5.** Superconductivity of B and C series samples. **(a)** Normalized *R-T* curves of pure B(P)SCCO (B1) and B(P)SCCO doped with 0.2 wt% $Y_2O_3$:$Sm^{3+}$ (B2), 0.2 wt% $Y_2O_3$ (B3), 0.2 wt% $Y_2O_3$:$Eu^{3+}$ (B4), 0.2 wt% $Y_2O_3$:$Eu^{3+}$+Ag (B5), and 0.3 wt% $Y_2O_3$:$Eu^{3+}$+Ag (B6). **(b)**, **(c)** The $\Delta T_{C,0}$ and $\Delta T_{C,on}$ of B2, B3, B4, B5, and B6 relative to B1. **(d)** Normalized *R-T* curves of pure B(P)SCCO (C1) and B(P)SCCO doped with 0.3 wt% $Y_2O_3$:$Sm^{3+}$ (C2), 0.3 wt% $Y_2O_3$ (C3), 0.3 wt% $Y_2O_3$:$Eu^{3+}$ (C4), 0.3 wt% $Y_2O_3$:$Eu^{3+}$+Ag (C5), 0.4 wt% $Y_2O_3$:$Eu^{3+}$ (C6), and 0.4 wt% $Y_2O_3$:$Eu^{3+}$+Ag (C7). **(e)**, **(f)** The $\Delta T_{C,0}$ and $\Delta T_{C,on}$ of C2, C3, C4, C5, C6, and C7 relative to C1.

The particle size of raw material S1 is 30 μm, and the optimal doping concentration of the inhomogeneous phase in the sample prepared by raw material S1 is 0.2%. At this time, compared with pure B(P)SCCO (A1), the $T_{C,0}$ and $T_{C,on}$ of B(P)SCCO doped with the $Y_2O_3$:$Eu^{3+}$ inhomogeneous phase increase by 0.5 and 0.7 K, respectively. Moreover, the $Y_2O_3$:$Eu^{3+}$+Ag inhomogeneous phase doped samples increase by 1 and 1.1 K, respectively. However, the particle size of raw material S3 is 5 μm, and the doping concentration of the inhomogeneous phase in the sample prepared by raw material S3 can be increased to 0.4%. At this time, compared with pure B(P)SCCO (C1), the $T_{C,0}$ and $T_{C,on}$ of B(P)SCCO doped with the $Y_2O_3$:$Eu^{3+}$ inhomogeneous phase increase by 2.5 and 4.6 K, and the $Y_2O_3$:$Eu^{3+}$+Ag inhomogeneous phase doped samples increase by 4 and 6.3 K, respectively. As the particle size of raw material decreases from 30 to 5 μm, the particle size of the B(P)SCCO superconducting matrix in the prepared sample gradually decreases, and the inhomogeneous phase doping concentration increases from 0.2% to 0.4%. At the same time, the increase of the inhomogeneous phase concentration enhances the $\Delta T_C$.

## 5. Conclusion

We designs a smart meta-superconductor B(P)SCCO with different matrix sizes. Three different series of samples were prepared by solid-state sintering using three kinds of particle sizes raw materials. The main phase of prepared samples is the high-temperature phase Bi2223 and contains a small amount of the low-temperature phase Bi2212. In addition, the microstructure consists of randomly distributed EL inhomogeneous phase and a plate-like B(P)SCCO matrix. $R$-$T$ tests found that the $T_C$ of $Y_2O_3$:$Eu^{3+}$+Ag or $Y_2O_3$:$Eu^{3+}$ luminescent inhomogeneous phase doping sample is higher than that of pure B(P)SCCO. Meanwhile, the $T_C$ of the $Y_2O_3$ or $Y_2O_3$:$Sm^{3+}$ non-luminescent dopant doping sample is lower than that of pure B(P)SCCO. As the particle size of raw material decreases from 30 to 5 μm, the particle size of the B(P)SCCO superconducting matrix decreases gradually, and the doping content of the $Y_2O_3$:$Eu^{3+}$+Ag and $Y_2O_3$:$Eu^{3+}$ inhomogeneous phases increases from 0.2% to 0.4%. Meanwhile, the growth of the inhomogeneous phase content further enhances the increase amplitude $\Delta T_C$. When the raw material particle size is 5 μm, the inhomogeneous phase added to the prepared sample can be increased to 0.4%. At this time, the $T_{C,0}$ and $T_{C,on}$ of $Y_2O_3$:$Eu^{3+}$+Ag doped sample are 4 and 6.3 K higher than those of pure B(P)SCCO, respectively.

**Author Contributions:** X.Z. conceived and led the project; H.C. and X.Z. designed the experiments; H.C., M.W., Y.Q. and Y.L. performed the experiments and characterized the samples; all authors discussed and analyzed the results; H.C. wrote the paper with input from all co-authors; X.Z. and H.C. discussed the results and revised the manuscript.

**Funding:** This work was supported by the National Natural Science Foundation of China for Distinguished Young Scholar under Grant No. 50025207.

**Date Avaliability Statement:** The date presented in this study are available on reasonable request from the corresponding author.

**Conflicts of Interest:** The authors declare no conflict of interest.